\documentclass[aps,pra,twocolumn,showpacs,preprintnumbers,amsmath,amssymb,citeautoscript,superscriptaddress,bibliography]{revtex4-2}
\usepackage{amsbsy,amssymb,amsmath,bm}
\usepackage{graphicx}
\usepackage{braket}
\usepackage{dsfont}
\usepackage{float}
\usepackage{textcomp}
\usepackage{color}
\usepackage[colorlinks=true,linkcolor=red]{hyperref}
\usepackage[sort&compress]{natbib}
\usepackage[normalem]{ulem}
\usepackage[shortlabels]{enumitem}
\graphicspath{{../Images/}}

\usepackage{float}
\usepackage[caption = false]{subfig}
\usepackage{amsmath}

\input{epsf}

\begin{document}
\title{
Linear Response for pseudo-Hermitian Hamiltonian Systems: Application to \textit{PT}-Symmetric Qubits
}
\author{L. Tetling}
\affiliation{ Institut f\"ur Theoretische Physik III, Ruhr-Universit\"at Bochum, Bochum 44801, Germany}

\author{M. V. Fistul}
\affiliation{ Institut f\"ur Theoretische Physik III, Ruhr-Universit\"at Bochum, Bochum 44801, Germany}

\author{Ilya M. Eremin}
\affiliation{ Institut f\"ur Theoretische Physik III, Ruhr-Universit\"at Bochum, Bochum 44801, Germany}

\date{\today}

\begin{abstract}
Motivated by the recent advances in modelling the \textit{pseudo-Hermitian Hamiltonian} (pHH) systems using superconducting qubits we analyze their quantum dynamics  subject to a small time-dependent perturbation. 
In particular, We develop the linear response theory formulation suitable for application to various pHH systems and compare it to the ones available in the literature. We derive analytical expressions for the generalized temporal quantum-mechanical correlation function $C(t)$ and the time-dependent dynamic susceptibility $\chi(t) \propto \text{Im} ~C(t)$.  We apply our results to two \textit{PT}-symmetric non-Hermitian quantum systems: a single qubit and two unbiased/biased qubits coupled by the exchange interaction. For both systems we obtain the eigenvalues and eigenfunctions of the Hamiltonian, identify \textit{PT}-symmetry unbroken and broken quantum phases and quantum phase transitions between them. The temporal oscillations of the dynamic susceptibility of the qubits polarization ($z$-projection of the total spin), $\chi(t)$, relate to {\it ac} induced transitions between different eigenstates and we analyze the dependencies of the oscillations frequency and the amplitude on the gain/loss parameter $\gamma$ and the interaction strength $g$. Studying the time dependence of $\chi(t)$ we observe different types of oscillations, i.e. undamped, heavily damped and amplified ones, related to the transitions between eigenstates with broken (unbroken) $PT$-symmetry. These predictions
can be verified in the microwave transmission experiments allowing controlled simulation of the pHH systems.  
\end{abstract}

\maketitle

\section{Introduction}\label{chap1}
Coherent quantum mechanics on the macroscopic scale is well established for isolated systems, and many fascinating effects like quantum beats and microwave induced Rabi oscillations \cite{kjaergaard2020superconducting}, precise manipulation of quantum bits (qubits) \cite{arute2019quantum}, maximally entangled Bell and Greenberger–Horne–Zeilinger (GHZ) states \cite{steffen2006measurement,yang2016entangling,shulga2021time}, collective quantum phases and phase transitions \cite{king2018observation,king2021scaling}, and non-equilibrium time-crystals \cite{yao2017discrete,choi2017observation,zhang2017observation} have been observed in optical, magnetic, semiconducting and superconducting  systems \cite{bruss2019quantum}. However, even a weak interaction of the quantum system with an environment results in unavoidable dissipation and decoherence leading to the relaxation of an excited quantum state population or the decay of quantum coherent oscillations on large times. 

A rapid development of quantum information technologies has allowed not just to fabricate quantum systems extremely weakly interacting with an environment, which is a necessary condition to observe the coherent quantum dynamics at larger times, but also to realize the opposite effect with a non-equilibrium growth of the population of specially chosen quantum states, {\it i.e}  the so-called states with a \textit{gain}. Consequently, a {\it loss} present in other parts of the system is then completely equalized by an induced gain and, as a result, the system can be described within the parity-time (\textit{PT})-symmetric non-Hermitian Hamiltonian, which belongs to the broader class of pseudo-hermitian Hamiltonins (pHH)\cite{mostafazadeh2002pseudo}. 
The dynamics governed by \textit{PT}-symmetric non-Hermitian Hamiltonians has been implemented in various one- and two dimensional photonic lattices \cite{ruter2010observation,el2018non,szameit2011p}, trapped ions and ultracold atoms \cite{ding2021experimental,li2019observation}, Bose-Einstein condensate \cite{cartarius2012model}, as well as superconducting \cite{naghiloo2019,dogra2021quantum} and nitrogen-vacancies qubits \cite{wu2019observation}. Here, the parity (\textit{P}) operator in arrays of interacting spins (qubits) is the direct product of local $\hat \sigma^{x}_i$ operators, while the transformation of $i \rightarrow -i$ determines the time-reversal symmetry.   

A theoretical analysis of the systems, whose quantum dynamics is described by an arbitrary non-Hermitian Hamiltonian, started long time ago \cite{dattoli1990non} but was boosted enormously by the seminal works of C. Bender with co-workers \cite{bender1998real,bender1999pt,bender2007making}. In particular, they have shown that the \textit{PT}-symmetric non-Hermitian Hamiltonian can exhibit a purely real eigenvalues spectrum, identifying the \textit{unbroken} \textit{PT}-symmetric quantum phase. At the same time as the gain/loss parameter varies there is also another regime of the Hamiltonian where the eigenvalues of the \textit{PT}-symmetric Hamiltonian become complex conjugate ones, signalling the \textit{broken} \textit{PT}-symmetric quantum phase, where the so-called exceptional point (line) determines the quantum phase transition between \textit{PT}-symmetric preserved and broken quantum phases. 

These unique quantum phases and transitions between them have been observed in overwhelming numbers of experimental studies \cite{ruter2010observation,el2018non,szameit2011p,ding2021experimental,li2019observation,naghiloo2019,dogra2021quantum}. More complex and intriguing physical phenomena like the flat bands \cite{leykam2017flat}, anyonic-parity-time symmetry \cite{arwas2022anyonic}, topological defects \cite{stegmaier2021topological,yuce2018pt} and topological phases \cite{liu2019second,Ghatak_2019,ju2019non,gneiting2022unraveling,Abbasi2022}, have been investigated in \textit{PT}-symmetric non-Hermitian Hamiltonian systems. Typically the dynamics of such systems was experimentally studied through the observation of specific time dependencies of the excited state population, i.e. the oscillating (\textit{PT}-symmetry preserved quantum phase) and decaying (\textit{PT}-symmetry broken quantum phases) ones. 

At the same another powerful method to measure the various transitions in Hamiltonian systems\cite{blais2007quantum,macha2014implementation,shulga2018magnetically,jung2014multistability} is the spectroscopic one in which the systems dynamics is probed by applying an external small time-dependent perturbation usually in the form of the electromagnetic field. For Hermitian systems the quantitative analysis of this spectroscopic method is based on the well-established Kubo linear response theory \cite{kubo1957statistical,dittrich1998quantum}. Therefore, natural questions arise: what is the linear response of a \textit{PT}-symmetric non-Hermitian Hamiltonian system to a low intensity external $ac$ electromagnetic field? What are the transitions that can be excited by a weak time-dependent perturbation in non-Hermitian systems? Previously the question on the linear response  was analyzed in various systems such as the non-Hermitian Dirac or Weyl Hamiltonians \cite{sticlet2022kubo,Kiss2022,KaiLi2022}, or the one- and two-dimensional Bose-Hubbard model \cite{Geier2021,pan2020non} assuming non-Hermitian/Hermitian types of perturbation \cite{sticlet2022kubo,Geier2021,Kiss2022} or complex non-Hermitian dissipation operator \cite{pan2020non}.

In this manuscript we discuss a generic \textit{linear response theory} for  \textit{PT}-symmetric non-Hermitian Hamiltonian systems subject to a small time-dependent Hermitian (physical) perturbation. Considering the \textit{PT}-symmetry Hamiltonian as a particular class of the pseudo-Hermitian Hamiltonian (pHH) \cite{mostafazadeh2002pseudo,mostafazadeh2,mostafazadeh3,mostafazadeh2020}, we provide a straightforward way to obtain the generalized temporal correlation function $C_{aa}(t)$ and the linear response function $\chi_{aa}(t)$ of the physical observable $a$ keeping in mind that the relevant experimental setup in which the coupling between a non-hermitian system and an external electromagnetic field occurs through a physical observable, i.e. a Hermitian operator. We apply our generic results to an analytical and numerical study of the quantum dynamics in two specific quantum systems: $PT$-symmetric single qubit and two interacting qubits. In particular, we obtain the dependence of eigenvalues $E_i$ on crucial parameters of the system, i.e. the gain/loss and the qubits interaction strength, identify various quantum phases and exceptional points (lines), and calculate the generalized temporal correlation function $C(t)$ and the linear response function $\chi(t)$ of the total polarization of qubits. The oscillations of $\chi(t)$ determine ac induced transitions between various states in \textit{PT}-symmetric non-Hermitian Hamiltonian systems.

The paper is organized as follows: In Section II we remind the main points of the quantitative description of the pHH system dynamics, introduce the pseudo-metric operator $\hat \eta$ and demonstrate a generic procedure how to calculate $\hat \eta $. In Section III the dynamics of pHH systems subject to a small time-dependent perturbation is studied, and the generic expressions for the time-dependent response function and the generalized temporal correlation function will be obtained. In next two Sections we calculate analytically and numerically the eigenvalues and eigenfunctions for two \textit{PT}-symmetric exemplary non-Hermitian quantum systems: a single qubit (Section IV) and two coupled qubits with the exchange type of interaction (Section V). Analyzing the observed oscillations of the imaginary part of the temporal correlation function of the qubit polarization, $\chi(t)$,  the dependencies of ac induced transitions between different states on the gain/loss parameter $\gamma$ and the interaction strength $g$, are obtained. Both \textit{PT}-symmetry unbroken and broken quantum phases are studied. 
We conclude with Section VI. 

\section{General description of 
pseudo-Hermitian Hamiltonian systems}\label{chap2}
Following to the seminal works \cite{bender1998real,bender2007making} and keeping in mind the application of the elaborated analysis to particular physical systems, i.e. interacting qubits, we review here a general  mathematical description of the \textit{pseudo-Hermitian} Hamiltonian  quantum dynamics \cite{mostafazadeh2002pseudo,mostafazadeh2,mostafazadeh3,mostafazadeh2020}.

Let us consider a system whose the dynamics is completely determined by pHH, i.e., the Hamiltonian $\hat H_s$ satisfies the following condition
\begin{align} \label{ConditionPH}
    \hat{\eta} \hat{H_s} = \hat{H}_s^{\dagger}\hat{\eta},
\end{align}
where $\hat{\eta}=\hat{\eta}^{\dagger}$ is a pseudo-metric Hermitian operator. Note, $\hat \eta$ is not uniquely defined and, in general, there are $N$ non-commuting operators $\eta$ for an $N$ -dimensional pHH \cite{bian2020conserved,agarwal2022conserved}. Moreover, the operator $\eta$ can also be time-dependent \cite{naghiloo2019,dogra2021quantum,wu2019observation}. The dynamics of such a system is characterized by the time-dependent wave function $\Psi(t)$ satisfying the dynamic (Schr\"odinger-like) equation 
\begin{align} \label{wavefunction}
   i\hbar \frac{d \Psi(t)}{dt} = \hat{H_s} \Psi(t).
\end{align}
As $\hat H_s$ does not depend explicitly on time there are two conserved quantities for a \textit{fixed} pseudo-metric operator $\eta$:  the total general norm $\langle \Psi (t) |\hat \eta|\Psi (t) \rangle$, which is equal to $1$ with the proper normalization of $\Psi$, and the effective Hamiltonian $\langle \Psi (t) |\hat \eta \hat H_s|\Psi (t) \rangle$. 

Since the arbitrary physical measurement has to result in the observation of real values, we define the quantum-mechanical averaging of the physical observable $a$ \footnote{Notice here, that in contrast to the PT-symmetric non-Hermitian optical systems the measured quantities in PT-symmetric non-Hermitian qubits systems are $\bar a(t)/(\langle \Psi(t)|\hat{\mathds{1}}|\Psi(t) \rangle) $ but not $\bar a(t)$. See also a short comment at the end of Sec. IV.} as
\begin{equation} \label{observble}
\bar a (t)=\langle \Psi(t)| \hat a | \Psi(t) \rangle =\langle \Psi(0)| e^{i\hat{H_s}^{\dagger} t/\hbar} \hat a e^{ -i\hat{H_s}t/\hbar}| \Psi(0) \rangle.
\end{equation}
Here, the Hermitian operator $\hat a$ associates with the physical observable $a$, and $\Psi(0)$ is the wave function of the initial state. In order to use the  Heisenberg representation and the corresponding equation of motion it is convenient to introduce the pseudo-Hermitian operator $\hat A$ related to the operator $\hat a$ of the physical observable $a$, as 
\begin{equation} \label{PsObservable}
\hat A= \hat \eta^{-1} \hat a.
\end{equation}
Since $\hat a$ is the Hermitian operator one finds that $\hat A$ satisfies Eq.(\ref{ConditionPH}). Throughout the text we will now use the lowercase for the Hermitian physical observable operators and the uppercase for the pseudo-Hermitian ones. For time-dependent operators $\hat A(t)$ the standard Heisenberg representation is valid
\begin{align}
\hat{A}(t) &= e^{i\hat{H}_s t/\hbar}\hat{A}e^{-\mathrm{i}\hat{H}_s t/\hbar}\label{eq: TDoperators}
\end{align}
and, therefore, Eq. (\ref{observble}) can be rewritten as
\begin{align}
\bar a(t)=\langle \Psi(t)| \hat a | \Psi(t) \rangle = \langle \Psi(0)| \hat \eta \hat{A}(t)|\Psi (0) \rangle. \label{eq: Averagedefinition}
\end{align}
Note, Eqs. (\ref{ConditionPH})-(\ref{eq: Averagedefinition}) present themselves a complete description of the dynamics of arbitrary systems characterized by a time independent pHH. However, to proceed further we need to derive an explicit expression for the pseudo-metric operator $\hat \eta$. Assuming the finite Hilbert space and that the Hamiltonian $\hat H_s$ is diagonalizable, one can see that $\hat{H}_s$ has a discrete spectrum $E_n$ and a complete, bounded bi-orthonormal set of eigenfunctions: $|R_n \rangle$ (right states) and $|L_n \rangle$ (left states), where the equations $\langle L_{m}|R_{n}\rangle = \delta_{mn}$ and $\sum_{n}|R_{n}\rangle\langle L_{n}| = \mathds{1}$ are satisfied. Here, $|R_n \rangle$ and $|L_n \rangle$ are the eigenfunctions of the Hamiltonian $\hat{H}_s$ and $\hat{H}_s^{\dagger}$, accordingly: $\hat{H}_s|R_{n}\rangle = E_{n}|R_{n}\rangle$, $\hat{H}_s^{\dagger}|L_{n}\rangle = E_{n}^{*}|L_{n}\rangle$. Notice that the generalization to the degenerate eigenvalues is straightforward. The eigenvalues $E_n$ of $\hat{H}_s$ are either real or emerge in complex conjugated pairs \cite{mostafazadeh2010}.

By making use of the eigenvalues $E_n$ and eigenfunctions $|L_n \rangle$ the pseudo-metric operator $\hat{\eta}$ is explicitly calculated as 
\begin{align}
    \hat{\eta} = \sum_{E_{n}\in\mathbb{R}}|L_{n}\rangle\langle L_{n}|+\sum_{E_{n_{+}}\in\mathbb{C}}\left(|L_{n_{+}}\rangle\langle L_{n_{-}}|+|L_{n_{-}}\rangle\langle L_{n_{+}}| \right),\label{eta-operator}
\end{align}
where the first sum contains all \textit{real} eigenvalues and the second sum  runs over \textit{complex} eigenvalues with positive imaginary part. 

\section{Linear response theory for pseudo-Hermitian Hamiltonian systems}
The quantitative description elaborated in the previous section is a good starting point to derive a general expression for the linear response of a non-Hermitian Hamiltonian system subject to an externally applied time-dependent force $f(t)$. In this case, the total time-dependent Hamiltonian is written as
\begin{equation}
\hat H_{tot}(t)= \hat H_s+f(t) \hat b, \label{total-Hamiltonian}
\end{equation}
where $\hat b$ is the Hermitian operator. This Hamiltonian allows one to adequately describe the relevant experimental spectroscopic setups where a studied pseudo-Hermitian system is coupled to the electromagnetic field probe.  The time-dependent operator of the observable $a$ satisfies the dynamic equation:
\begin{equation}
i\hbar \frac{d \hat a }{dt}= \hat a \hat H_{tot}- \hat{H}^\dagger_{tot}\hat{a}´=  \hat a \hat H_{s}- \hat{H}^\dagger_{s}\hat{a}+f(t)[\hat a (t), \hat b].\label{response:dynequation1}
\end{equation}
By making use of the relation (\ref{PsObservable}) between the operators $\hat a$ and $\hat A$, and Eq. (\ref{eq: TDoperators}), the solution of Eq. (\ref{response:dynequation1}) is obtained in zeroth order of perturbation series over a small function $f(t)$  as 
\begin{equation}
\hat a_0(t)=\eta \hat A(t). \label{response:solutioinn0}
\end{equation}
Correspondingly, in the first order of perturbation series we write the solution as
\begin{equation}
\hat a_1(t)=-\frac{i}{\hbar}\int_0^t ds f(s) [\hat a_0(t-s),\hat b] \label{response:solution1}
\end{equation}
and defining the response function $\chi_{ab}(t)$ as 
$\bar a_1(t)=\langle \hat a_1(t) \rangle =\int_0^t ds \chi_{ab}(t-s)f(s)$ we obtain the response function for a pHH system
\begin{equation}
\chi_{ab}(t)=-\frac{i}{\hbar} \langle [\hat \eta \hat A(t), \hat \eta \hat B(0) ]\rangle_0 \Theta(t). \label{response_function1}
\end{equation}
Here, the step function $\Theta(t)$ is introduced to establish the causality, and the averaging $\langle...\rangle_0$ occurs over the wave function $\Psi(0)$ of the initial state. 
As $\hat b=\hat a$ that we will assume for the rest of the paper, the response function $\chi_{aa}(t)$ is obtained as
\begin{equation}
\chi_{aa}(t)=-\frac{i}{\hbar} \langle [\hat \eta \hat A(t), \hat \eta \hat A(0) ]\rangle_0 \Theta(t),\label{response_function1}
\end{equation}
where $\hat A(0)$ and $\hat A(t)$ are determined by Eqs. (\ref{PsObservable}) and (\ref{eq: TDoperators}), respectively.
The dynamic response of a pHH system to an external time-dependent perturbation can be also characterized by the generalized temporal correlation function $C_{aa}(t)$ written as
\begin{equation}
C_{aa}(t)=\langle (\hat \eta \hat{A}(t))\cdot (\hat \eta \hat{A}(0)) \rangle_0  \label{eq: Correlationfunction}
\end{equation}
and $\chi_{aa}(t)=(2/\hbar)\text{Im} C_{aa}(t)\Theta(t)$.

Note, Eq. (\ref{response_function1}) allows one to obtain interesting generic properties of the linear response of pHH systems. By inserting all possible intermediate states in Eq.(\ref{response_function1}), for systems with a \textit{PT-symmetry broken ground state}  we can write explicitly:
\begin{eqnarray}
\chi(t)& = & -i\theta (t) \sum_n ( <R_0|\hat \eta \hat A|R_n> \cdot <R_n|\hat \eta \hat A|R_0>  ) \times \nonumber \\
&& \left [e^{\frac{i}{\hbar}(E_1-E_n)t}-  e^{\frac{i}{\hbar}(E_n^{\star}-E_0)t} \right],  \label{eq: response-expl}
\end{eqnarray}
where the energies $E_{0}$ and $E_1$ pertain to the PT-symmetry broken ground state, {\it i.e.} $E_0=a-ib$ and $E_1=E_0^{\star}$. Thus,  for a single qubit with gain and loss, where $n$ takes the values $0,1$ only, we obtain $\chi(t)=0$ in the PT-symmetry broken regime. Therefore, a weak ac electromagnetic field \textit{cannot} excite the transitions between the eigenstates with complex conjugate eigenvalues, which is also a physically expected result \cite{sticlet2022kubo}. Also, for a more complex system where various combinations of PT-symmetry preserved (broken) states can be realized, Eq. (\ref{eq: response-expl}) demonstrates the presence of damped oscillations provided  by the transitions between the PT-symmetry broken state,  $n=0$, and PT-symmetry preserved excited states.

\begin{figure}
\centering
\includegraphics[width=3.5in,angle=0]{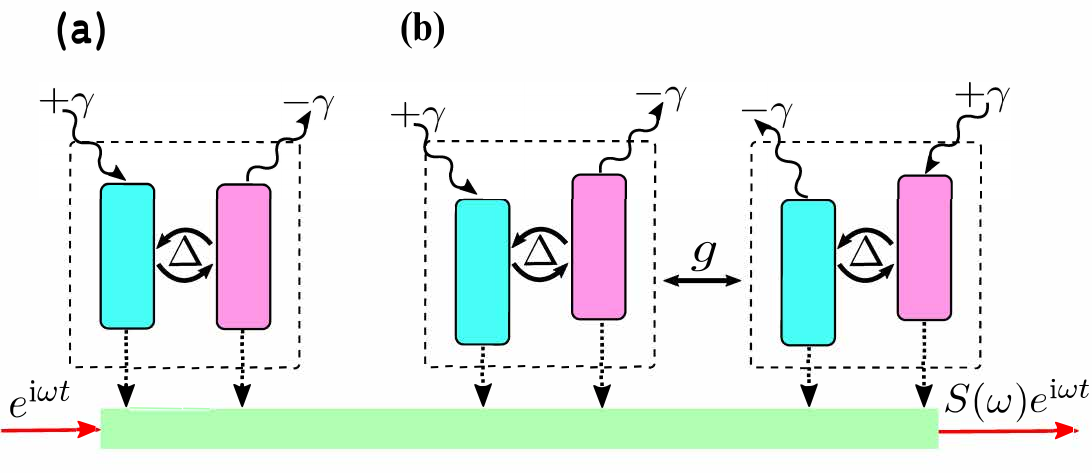}

\caption{ 
Schematic representation of PT-symmetric (a) single unbiased qubit and (b) two interacting biased qubits. Here, $\Delta$ refers to the off-diagonal coupling between the eigenstates, $\gamma$ is the processes of gain/loss (a staggered gain/loss) and $g$ is an exchange interaction between the qubits. A low-dissipative waveguide couples weakly to qubits and $S(\omega)$ refers the electromagnetic waves transmission coefficient. 
 }
\label{Pic.1}
\end{figure}

In next two Sections we now present a detailed analysis of the energy spectrum, phase diagrams and the response function for two exemplary models from the realm of qubits: \textit{PT}-symmetric single unbiased qubit and a \textit{PT}-symmetric array of interacting unbiased/biased qubits. The dynamics of these qubits systems is determined by the \textit{PT}-symmetric non-Hermitian Hamiltonians written in the spin representation as %
\begin{equation}
    \hat{H}_{SQ} = \Delta \hat \sigma^x+i\gamma \hat \sigma^z
\label{single_qubit}
\end{equation}
for a single unbiased qubit and
\begin{align}
    \hat{H}_{QA} &= \sum_{i=1}^M \left[\frac{\Delta}{2}\hat{\sigma}_{i}^{x}+\frac{\epsilon}{2}\hat{\sigma}_{i}^{z}+(-1)^{i}\,\mathrm{i}\gamma \hat{\sigma}_{i}^{z}\right]\nonumber\\
    &+g\sum_{ij}(\hat{\sigma}_{i}^{+}\hat{\sigma}_{j}^{-}+\hat{\sigma}_{i}^{-}\hat{\sigma}_{j}^{+}).\label{interacting_qubits}
\end{align}
for an array of $M$ interacting qubits. In the latter case the qubits exchange interaction of the strength $g$ is assumed. Here, $\hat \sigma_i^{x,y,z}$ are the corresponding Pauli matrices, $\Delta$, $\epsilon$ and $\gamma$ are of-diagonal and diagonal matrix elements, gain/loss parameter of individual qubits, respectively. For the sake of simplicity we assume that these parameters are identical for all qubits. The two setups are also shown schematically in Fig. \ref{Pic.1}. Note that for the larger size qubits system the numerical diagonalization may require an application of the more sophisticated algorithms.\cite{Lubasch2013,Lubasch2017} 

A weak coupling of the qubits system to a low-dissipative waveguide (shown at the bottom of Fig. \ref{Pic.1}) allows to \textit{directly} obtain $\chi(\omega)$, i.e., the Fourier transform of $\chi(t)$, by measuring of the singularities of the electromagnetic waves transmission coefficient \cite{blais2007quantum,macha2014implementation,shulga2018magnetically,jung2014multistability,fistul2022quantum,navez2022quantum}.

\section{Quantum dynamics of a \textit{PT}-symmetric unbiased qubit
}\label{chap4}
\begin{figure}
\centering
\includegraphics[width=2.4in,angle=0]{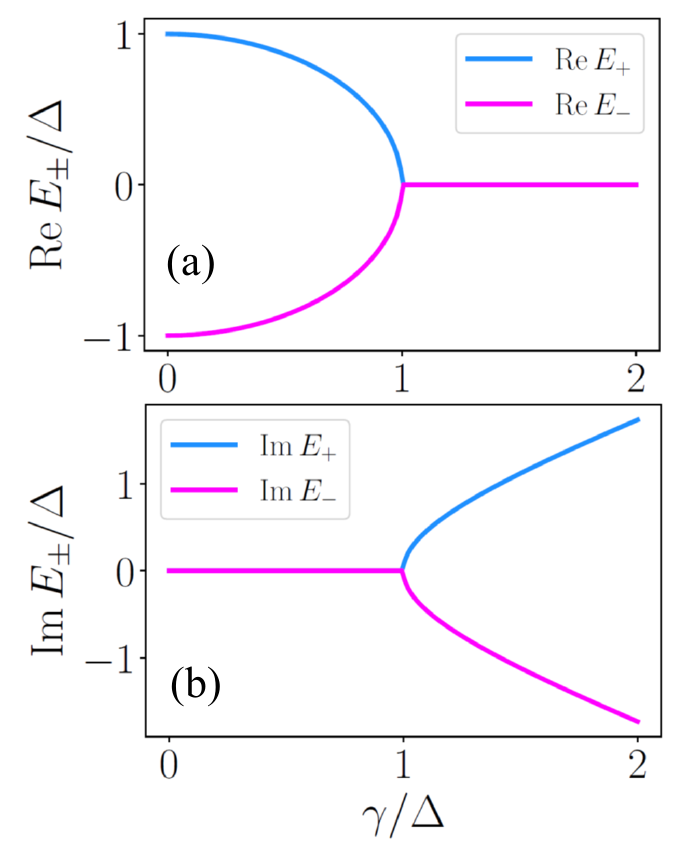}
\caption{ 
The dependencies of real (a) and imaginary (b) parts of eigenvalues $E_{\pm}$ on the gain/loss parameter $\gamma/\Delta$ demonstrating the \textit{PT}-symmetry preserved 
and broken 
quantum phases. The exceptional point is at $\gamma/\Delta=1$. 
 }
\label{Pic.2}
\end{figure}

The Hamiltonian (\ref{single_qubit}) of a single unbiased qubit is invariant under the combined action of operators $\hat{P} = \hat{\sigma}^{x}$ and $\hat{T} = \mathcal{K}\mathds{1}$ where time-reversal acts as complex conjugation. The eigenvalues of (\ref{single_qubit}) can be readily obtained as
\begin{align}
    E_{\pm} = \pm \sqrt{\Delta^{2}-\gamma^{2}}\label{SQ-eigenvalues}
\end{align}
and their dependencies on the parameter $\gamma/\Delta$ 
are presented in Fig. \ref{Pic.2}. For $\gamma < \Delta$ the eigenvalues are real and a qubit is in the \textit{PT}-symmetry \textit{preserved (unbroken)} regime, whereas for $\gamma > \Delta$ the eigenvalues $E_\pm$ are conjugated ones with the zero real part and a single qubit is in the \textit{PT}-symmetry \textit{broken} phase. The condition $\gamma =\Delta$ determines the exceptional point dividing the \textit{PT}-symmetry unbroken and broken quantum phases (see Fig. \ref{Pic.2}). Next we consider the linear response of the systems in the \textit{PT}-symmetry preserved and broken regimes.

\subsection{\textit{PT}-symmetry preserved regime}
By making use of the generic results obtained in Sections II  we derive the bi-orthogonal eigenvectors of the Hamiltonian (\ref{single_qubit}) as 
\begin{align}
    |R_{\pm}\rangle = \frac{\mathcal{N}_{\pm}}{\sqrt{2}}\begin{pmatrix}1 \\ \frac{\pm E-\mathrm{i}\gamma}{\Delta} \end{pmatrix}, \ \ |L_{\pm}\rangle = \frac{1}{\sqrt{2}}\begin{pmatrix}1 \\ \frac{\pm E+\mathrm{i}\gamma}{\Delta} \end{pmatrix}, \label{SQ-eigenfunctions1}
\end{align}
where the normalization coefficients are $\mathcal{N}_{\pm} = \pm\frac{\Delta^{2}}{ E \left(\pm E-\mathrm{i}\gamma \right)}$ and $E=\sqrt{\Delta^2-\gamma^2}$. One can verify that $\langle L_{m}|R_{n}\rangle = \delta_{mn}$ with $m, n = \pm$ and $\sum_{n = \pm}|R_{n}\rangle\langle L_{n}| = \mathds{1}$. Using Eq.(\ref{eta-operator}) the pseudo-metric operator $\hat \eta$ is then given by
\begin{align}
    \hat{\eta} &= |L_{+}\rangle\langle L_{+}|+|L_{\text{-}}\rangle\langle L_{\text{-}}|=\nonumber\\
    &= \begin{pmatrix}1 & \text{-}\mathrm{i}\frac{\gamma}{\Delta} \\ \mathrm{i}\frac{\gamma}{\Delta} & 1 \end{pmatrix}.\label{SQ-etaoperator}
\end{align}
Taking into account this expression for $\hat \eta$ we proceed with the derivation of the time-dependent qubit polarization ($z$-component of the spin), i.e. $\bar \sigma^z(t)$. To make that we obtain the pseudo-Hermitian operator $\hat \Sigma^z$ related to the relevant observable $\sigma^z$ as 
\begin{align}
    \hat{\Sigma}^z &= \hat \eta^{-1} \sigma^z = \nonumber\\
    &= \frac{\Delta^2}{\Delta^2-\gamma^2}\begin{pmatrix}1 & \text{-}\mathrm{i}\frac{\gamma}{\Delta} \\ \text{-}\mathrm{i}\frac{\gamma}{\Delta} & -1 \end{pmatrix}.\label{SQ-Sigma}
\end{align}
It is evident that in the limit of $\gamma \rightarrow 0$ the pseudo-Hermitian operator $\hat{\Sigma}^{z}$ becomes the Hermitian operator $\hat{\sigma}^{z}$. 

We remind that the explicit expression for the time-dependent qubits polarization $\bar \sigma^z(t)$ depends on the initial state $\Psi(0)$. Thus, the results become particularly transparent if the initial state is chosen in the basis of $z$-projections of spin, $|\uparrow\rangle = \begin{pmatrix}1 & 0 \end{pmatrix}^{\mathrm{T}}$. Such initial state presented in the basis of eigenstates of the Hamiltonian $\hat H_{SQ}$, is given by  $|\uparrow\rangle = \frac{1}{\sqrt{2}}(|R_{+}\rangle+|R_{-}\rangle)$.

Straightforward but somewhat lengthy calculations (see Eqs. (\ref{eqa3}-\ref{eqa6}) of the Appendix A) allow to derive the non-zero matrix elements of the operator $\hat \eta \hat \Sigma^z(t)$ as
\begin{align}
    \langle R_{-}|\hat \eta \hat{\Sigma}^{z}(t)|R_{+}\rangle = \frac{\Delta^2 }{E(E-\mathrm{i}\gamma)}e^{-2\mathrm{i}Et/\hbar}\label{ME-Sigma}
\end{align}
and $ \langle R_{+}|\hat \eta \hat{\Sigma}^{z}(t)|R_{-}\rangle =\langle R_{-}|\hat \eta \hat{\Sigma}^{z}(t)|R_{+}\rangle^{*}$. By making use of Eqs. (\ref{eq: Averagedefinition}) and (\ref{ME-Sigma}) we obtain in the \textit{PT}-symmetry preserved regime
\begin{equation}
    \bar \sigma_{ubr}^z(t)=\langle \Psi(0)|\eta \hat{\Sigma}^{z}(t)|\Psi(0)\rangle = \mathrm{cos}(2Et/\hbar)+\frac{\gamma}{E}\mathrm{sin}(2Et/\hbar).\label{TDpolarization}
\end{equation}
Not surprisingly we obtain that the quantum dynamics of a single qubit demonstrates \textit{undamped} oscillations with the frequency $\omega=\sqrt{\Delta^2-\gamma^2}/\hbar$ in the \textit{PT}-symmetry preserved regime. 

Next we turn to the analysis of the temporal correlation function of the qubits polarization that can be written as
\begin{equation}
C_{ubr}(t)=\langle \hat \eta \hat{\Sigma}^z (t) \cdot \hat \eta \hat{\Sigma}^z \rangle_0. \label{eq: CorrelationfunctionSQ}
\end{equation}
In order to study the ac induced transitions we choose  $|R_- \rangle$ as the initial state, that is the ground state of the Hamiltonian $\hat H_{SQ}$. Using matrix elements of $\hat \Sigma^z$ (see Eqs. (\ref{eqa3})-(\ref{eqa6}))  we obtain the temporal correlation function $C_{ubr}(t)$
\begin{equation}
C_{ubr}(t)=\frac{\Delta^2}{E^2}\exp(-2iEt/\hbar). \label{eq: CorrelationfunctionSQ-Fin}
\end{equation}

The response function of a single qubit in the \textit{PT}-symmetry preserved regime is determined by the imaginary part of $C_{ubr}(t)$ as
\begin{align}
    \chi_{ubr}(t) =\frac{2}{\hbar}\theta(t) \text{Im} C_{ubr}(t)= 2\frac{\Delta^2}{E^2}\theta(t)\mathrm{sin}(2Et/\hbar).\label{eq:responseSQ-ubr}
\end{align}
To conclude this subsection we notice that the Fourier transform of the $\text{Im} ~C_{ubr}(t)$, i.e. $\chi_{ubr}(\omega)$, determines the dynamic susceptibility of the system. In the \textit{PT}-symmetry preserved regime the $\chi_{ubr}(\omega)$ displays a singularity at the frequency $2\sqrt{\Delta^2-\gamma^2}$. This singularity is related to $ac$-induced transitions between the pHH eigenstates.  

\subsection{\textit{PT}-symmetry broken regime}
In the \textit{PT}-symmetry broken regime $\gamma > \Delta$ and the eigenvalues are purely imaginary, {\it i.e.} $E_\pm=\pm i \tilde E$, where $\tilde E =\sqrt{\gamma^2-\Delta^2}$. The eigenvectors are given by 
\begin{align}
    |R_{\pm}\rangle = \frac{\mathcal{N}_{\pm}}{\sqrt{2}}\begin{pmatrix}1 \\ \frac{\pm \mathrm{i}\tilde{E}-\mathrm{i}\gamma}{\Delta} \end{pmatrix}, \ \ |L_{\pm}\rangle = \frac{1}{\sqrt{2}}\begin{pmatrix}1 \\ \frac{\mp \mathrm{i}\tilde{E}+\mathrm{i}\gamma}{\Delta} \end{pmatrix}, \label{eq:eigenvectorsBR}
\end{align}
where the normalization coefficients are $\mathcal{N}_{\pm} = -\Delta^{2}/[\tilde{E}(\tilde{E}\mp\gamma)]$. The orthogonality of the eigenstates is $\langle R_n |L_m \rangle=\delta_{n, m}$, where $n,m=\pm$. 

In the \textit{PT}-symmetry broken regime the pseudo-metric operator $\hat \eta=|L_{+}\rangle\langle L_{-}|+|L_{-}\rangle\langle L_{+}|$ is still given by Eq. (\ref{SQ-etaoperator}), and also the pseudo-Hermitian operator $\hat \Sigma^z$ is determined by Eq. (\ref{SQ-Sigma}). 
By making use of the results from the Appendix B (see Eqs. (\ref{eqb3})-(\ref{eqb8})), and choosing the initial state as $|\uparrow \rangle = (|R_+\rangle + |R_-\rangle)/\sqrt{2}$ we obtain the time-dependent qubits polarization in the \textit{PT}-symmetry broken regime as 
\begin{equation}
    \bar \sigma_{br}^z(t)=\langle \Psi(0)|\hat \sigma^z(t)|\Psi(0)\rangle = \mathrm{cosh}(2\tilde{E}t/\hbar)+\frac{\gamma}{E}\mathrm{sinh}(2\tilde{E}t/\hbar).\label{TDpolarizationBR}
\end{equation}
Thus, one can see that in the \textit{PT}-symmetry broken regime the oscillations of the qubit polarization are absent and not just damped, which is a consequence of the eigenenergies being purely imaginary. 

In the \textit{PT}-symmetry broken regime the temporal correlation function of the qubit polarization $C_{{br}}(t)$, calculated for the initial state $|R_-\rangle $, contains only the real part 
\begin{equation}
C_{{br}}(t)=\frac{\Delta^2}{\tilde{E}^2}\exp(-2\tilde{E}t/\hbar). \label{eq: CorrelationfunctionSQBR-Fin}
\end{equation}
and therefore, the linear response function $\chi_{{br}}(t)$ is \textit{zero} in the \textit{PT}-symmetry broken regime in correspondence to the general equation (\ref{eq: response-expl}). 

To conclude this subsection we notice that Eqs. (\ref{TDpolarization}) and (\ref{TDpolarizationBR}) for the time-dependent qubits polarization in both preserved and broken regimes, have been confirmed in experiments with superconducting \cite{naghiloo2019,dogra2021quantum}  and nitrogen-vacancies \cite{wu2019observation} qubits, where the quantity $\bar \sigma^z(t)/(\langle \Psi(t)|\hat{\mathds{1}}|\Psi(t) \rangle) $ was measured. Thus our approach could be directly applied there.

\section{Quantum dynamics of \textit{PT}-symmetric two interacting qubits}\label{chap5}
The quantum dynamics of two interacting \textit{ PT}-symmetric qubits, shown in Fig. \ref{Pic.1}(b), is governed by the Hamiltonian (\ref{interacting_qubits}) with $M=2$. We stress here that in the presence of the exchange type of interaction between the qubits, and opposite signs of the gain/loss parameter for different qubits, i.e. a staggered gain/loss, the \textit{PT}-symmetry of the Hamiltonian (\ref{interacting_qubits}) is preserved even in the biased regime, $\epsilon \neq 0$. However, the $PT$-symmetry in the system of interacting qubits with an arbitrary bias $\epsilon$ can only be achieved if the total number of qubits is even. Indeed, for an $M$-qubit configuration we define parity and time-reversal as
\begin{align}
    \hat{P}\hat{\sigma}^{\alpha}_{j}\hat{P}^{-1} = \hat{\sigma}^{\alpha}_{N+1-j}, \ \ \hat{T}\mathrm{i}\hat{T}^{-1} = -\mathrm{i}.\label{eq32}
\end{align}
and the operator $\hat P \hat T$ commutes with the pHH (\ref{interacting_qubits}). 
The disorder in qubits parameters will also break the \textit{PT}-symmetry of the Hamiltonian (\ref{interacting_qubits}) but some signatures of the symmetry can still be measured even in a system that inherently breaks $PT$-symmetry \cite{naghiloo2019}. 

Our strategy is to numerically calculate the eigenvalues $E_i$ and eigenvectors $|R_i\rangle$ ($|L_i\rangle$) in order to compute the time-dependent linear response of the total qubits polarization for different values of the system parameters like $\Delta$, $\epsilon$, $\gamma$ and $g$. We identify various unbroken and broken \textit{PT}-symmetry quantum phases and corresponding quantum phase transitions for two distinguishable cases, i.e., unbiased and biased interacting qubits. 

\subsection{Two unbiased interacting qubits, $\epsilon=0$}

The quantum dynamics of two interacting unbiased qubits is determined by the pHH written explicitly as
\begin{equation} 
\hat H_{2Q}=  \frac{\Delta}{2}\hat{\sigma}_{1}^{x}+\frac{\Delta}{2}\hat{\sigma}_{2}^{x}-i\gamma \hat{\sigma}_{1}^{z}+i\gamma \hat{\sigma}_{2}^{z} +g(\hat{\sigma}_{1}^{+}\hat{\sigma}_{2}^{-}+\hat{\sigma}_{1}^{-}\hat{\sigma}_{2}^{+}).\label{twoqubits}
\end{equation}

The eigenvalues $E_i$, where $i=1,..4$, of $\hat H_{2Q}$  are obtained as solutions of the secular equation $P_0(E)=0$
\begin{equation}
P_0(E)=E^{4}+(\gamma^{2}-\Delta^{2}-g^{2})E^{2}-g\Delta^{2}E\label{solutionEV_eps0}.
\end{equation}

\begin{widetext}

\begin{figure}[t]
\includegraphics[width=\linewidth,angle=0]{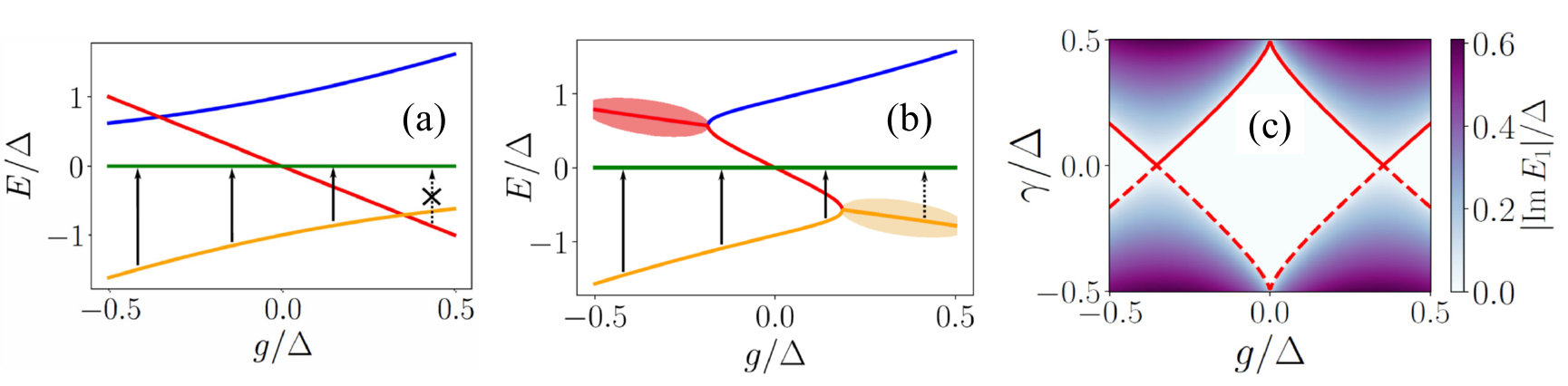}
\caption{The eigenvalues of two interacting PT-symmetric unbiased qubits as a function of the interaction strength $g/\Delta$ for  $\gamma=0$ (a) and $\gamma/\Delta=0.2$ (b). The eigenvalues having an imaginary part are shown by shaded area. The arrows indicate the ac induced transitions between energy levels that can be observed in the linear response $\chi_{2Q}(g;t)$. c) shows a complete phase diagram  demonstrating the presence of preserved (white area) and broken (blue area) \textit{PT}-symmetry quantum phases. The lines of exceptional points determining the transitions between quantum phases are indicated by red lines.}
\label{fig:spectrum_PD_e_0}
\end{figure}

\end{widetext}

Observe that independently of  the qubits parameters one of the eigenvalues is always zero, $E=0$. The other three eigenvalues strongly vary with $g$ and $\gamma$. The typical dependencies of eigenvalues $E_{1-4}$ on the strength of the interaction $g$ are shown in Fig. \ref{fig:spectrum_PD_e_0}a,b for two values of $\gamma$: $\gamma=0$ (the Hermitian quantum regime) and $\gamma/\Delta=0.2$ (the pseudo-Hermitian quantum regime).

For $\gamma \neq 0$ once the interaction strength $|g|$ overcomes the critical values the \textit{two} \textit{PT}-symmetry broken quantum phases (indicated in Fig. \ref{fig:spectrum_PD_e_0}b by shaded areas) are realized. In these quantum phases some eigenvalues are complex conjugated ones. Moreover, for negative ( positive) values of $g$  the complex eigenvalues were obtained for the "ground" (excited) state, and the quantum phase transitions (QPTs) between the PT symmetry preserved and PT-symmetry broken quantum phases can be found, which is similar to the well-known QPTs observed in Hermitian systems of interacting spins (see the regions of $|g| \geq 0.4$ in Fig. \ref{fig:spectrum_PD_e_0}a). The complete phase diagram $\gamma/\Delta-g/\Delta$ demonstrating the presence of preserved and broken \textit{PT}-symmetry quantum phases and the lines of exceptional points dividing these quantum phases is shown in Fig. \ref{fig:spectrum_PD_e_0}c. Comparing the Figs. \ref{fig:spectrum_PD_e_0}a and \ref{fig:spectrum_PD_e_0}b one finds that the critical strengths determining the QPTs between broken and unbroken quantum phases shift to lower values as $\gamma$ increases . 

\begin{figure}[h]
\centering
\includegraphics[width=3.5in,angle=0]{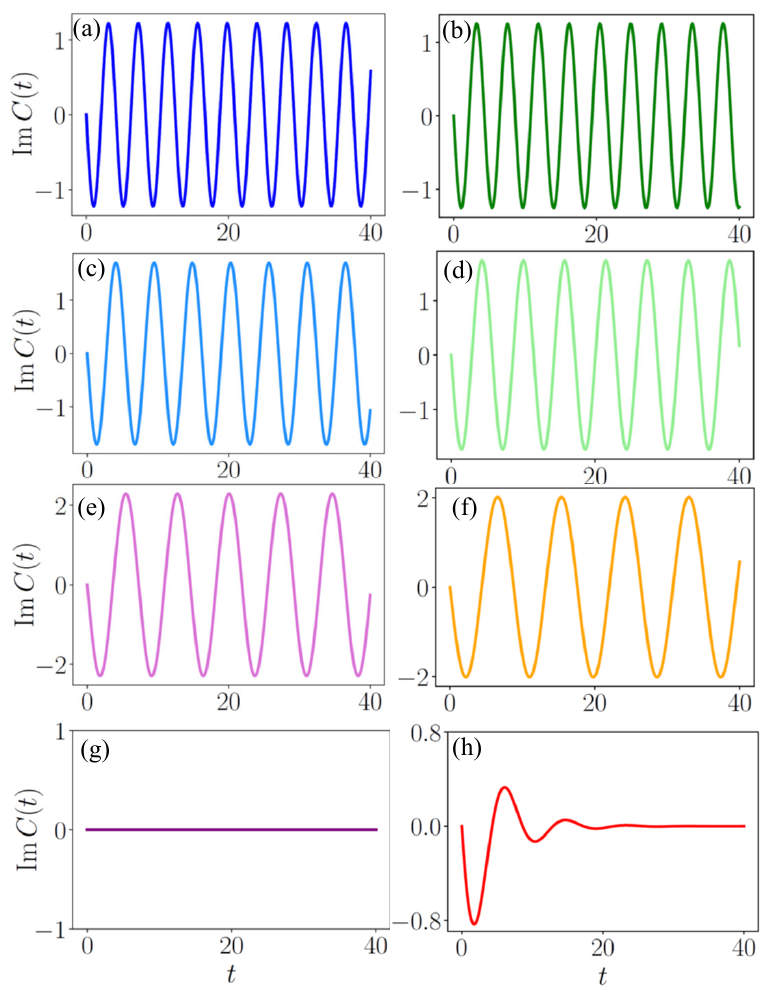}
\caption{Time-dependent linear response $\chi_{2Q}(t)$ of the total polarization of two interacting unbiased qubits ($\epsilon=0$) for two different values of the gain/loss parameter $\gamma$, i.e. $\gamma=0$ (left panels) and $\gamma/\Delta=0.2$ (right panels). The interaction strength $g$ was chosen as  $g/\Delta=-0.42$ (a)-(b); $g/\Delta=-0.15$ (c)-(d); $g/\Delta=0.15$ (e)-(f); $g/\Delta=0.42$(g)-(h).}
\label{fig:linear-response_e_0_G0}
\end{figure}

Fixing the gain/loss parameter $\gamma/\Delta$ and choosing the interaction strength $g$ above and below of the critical values determining the QPTs between unbroken and broken quantum phases, making use of obtained eigenvalues $E_i$ and eigenfunctions $|R_i\rangle$ ($|L_i\rangle$) we numerically calculate the pseudo-metric operator $\hat \eta_{2Q}$ and the pseudo-Hermitian operator of the total polarization $\hat \Sigma_{2Q}=\hat \eta_{2Q}^{-1} (\hat \sigma_1^{z}+\hat \sigma_2^{z})$. In our analysis we use the minimum energy state of the Hamiltonian (\ref{twoqubits}) for the initial state $\Psi(0)$. Using further the general results (\ref{response_function1}), (\ref{eq: Correlationfunction}) we compute the linear response of the total qubits polarization for two interacting qubits $\chi_{2Q}(g;t)$. in Fig. \ref{fig:linear-response_e_0_G0} we show the time-dependencies of $\chi_{2Q}(g;t)$ for several particular values of the interaction strength $g$.

Comparing the left and the right panels of Fig. \ref{fig:linear-response_e_0_G0} we register some interesting features of the linear response in the pHH system. For example in the presence of the \textit{PT}-symmetric gain/loss  $\gamma$ the linear response $\chi_{2Q}(t)$ of a \textit{PT}-symmetry unbroken quantum phase demonstrates undamped quantum oscillations (see Fig. \ref{fig:linear-response_e_0_G0}d,f). These oscillations are fingerprints of ac induced transitions between the "ground" and excited states indicated by solid arrows in Fig. \ref{fig:spectrum_PD_e_0}a,b. The frequencies of such oscillations decrease substantially with respect to the  Hermitian case (compare the panels (c,e) and (d,f) in Fig. \ref{fig:linear-response_e_0_G0}). Notice here that the undamped oscillations also present for negative values of $g<g^{(-)}_{cr}$ in the \textit{PT}-symmetry broken quantum phase  (see Fig. \ref{fig:linear-response_e_0_G0}b) in which the transitions to the eigenstates with complex conjugated eigenvalues are forbidden. A most spectacular difference between the Hermitian and \textit{PT}-symmetric non-Hermitian two interacting qubits system is obtained for large positive values of $g>g^{(+)}_{cr}$, e.g. for $g=0.42$. Indeed, if for $\gamma=0$ all transitions are completely forbidden (see Fig. \ref{fig:linear-response_e_0_G0}g), the linear response of the \textit{PT}-symmetry broken phase shows highly damped oscillations (see Fig. \ref{fig:linear-response_e_0_G0}h). These oscillations correspond to the ac induced transition indicated by dashed arrow in Fig. \ref{fig:spectrum_PD_e_0}b. Notice here, that the presence of such highly damped oscillations is the direct consequence of  Eq. (\ref{eq: response-expl}).

\subsection{Two biased interacting qubits, $\epsilon \neq 0$}

The quantum dynamics of two interacting qubits with an arbitrary bias $\epsilon$ is determined by the pHH (\ref{interacting_qubits}) with $M=2$. The eigenvalues $E_i$, where $i=1,..4$, of (\ref{interacting_qubits}) are obtained as a solution of the secular equation $P_\epsilon (E)=0$ with
\begin{align}
    P_\epsilon(E) &= E^{4}+(\gamma^{2}-\Delta^{2}-g^{2}-\epsilon^{2})E^{2}-g\Delta^{2}E\nonumber\\
    &+\epsilon^{2}(g^{2}-\gamma^{2}).\label{solutionEV_eps}
\end{align}

\begin{widetext}

\begin{figure}[h]
\centering
\includegraphics[width=\linewidth,angle=0]{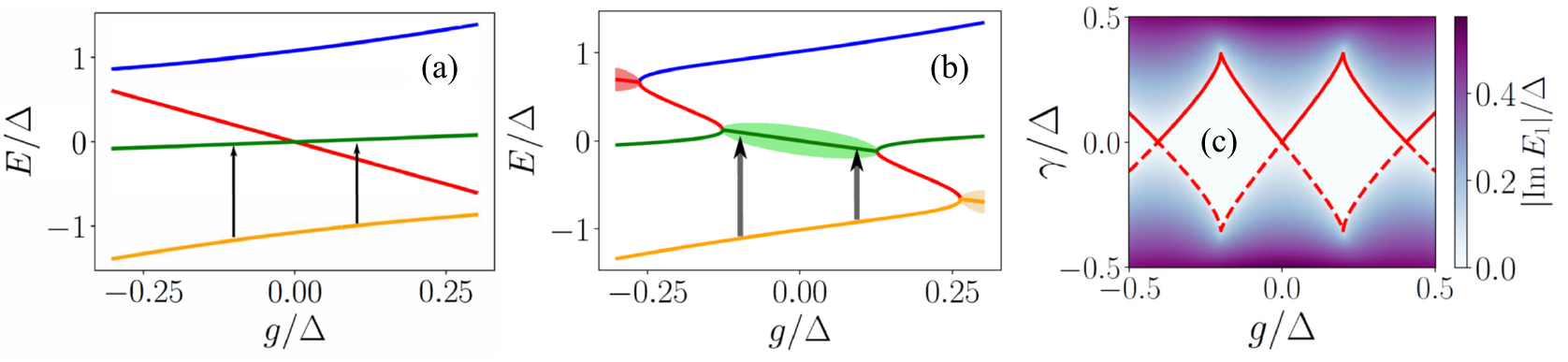}
\caption{The eigenvalues of two interacting PT-symmetric  qubits with the bias $\epsilon/\delta=0.2$ as a function of the interaction strength $g/\Delta$ for $\gamma/\Delta=0$ (a) and $\gamma/\Delta=0.2$ (b). The eigenvalues having an imaginary part are shown by shaded area. (c) shows the complete phase diagram demonstrating the presence of unbroken (white area) and broken (blue area) \textit{PT}-symmetry quantum phases. The exceptional points determining the transitions between quantum phases are indicated by red lines.}
\label{fig:spectrum_PD_e_02}
\end{figure}

\end{widetext}

The typical dependencies of eigenvalues $E_i$ on the strength of interaction $g$ for biased qubits are shown in Fig. \ref{fig:spectrum_PD_e_02}a,b for $\gamma=0$ and $\gamma/\Delta=0.2$, respectively. Similar to the case of unbiased qubits the two \textit{PT}-symmetry broken quantum phases occur for large positive (negative) values of $g$. However, in the presence of a bias, $\epsilon \neq 0$ and for low values of interaction strength $g$ we obtain an additional \textit{PT}-symmetry broken quantum phase (see Fig. \ref{fig:spectrum_PD_e_02}b green shaded area). The complete phase diagram $\gamma/\Delta-g/\Delta$ for biased interacting qubits is presented in Fig. \ref{fig:spectrum_PD_e_02}c.

\begin{figure}
\centering
\includegraphics[width=3.5in,angle=0]{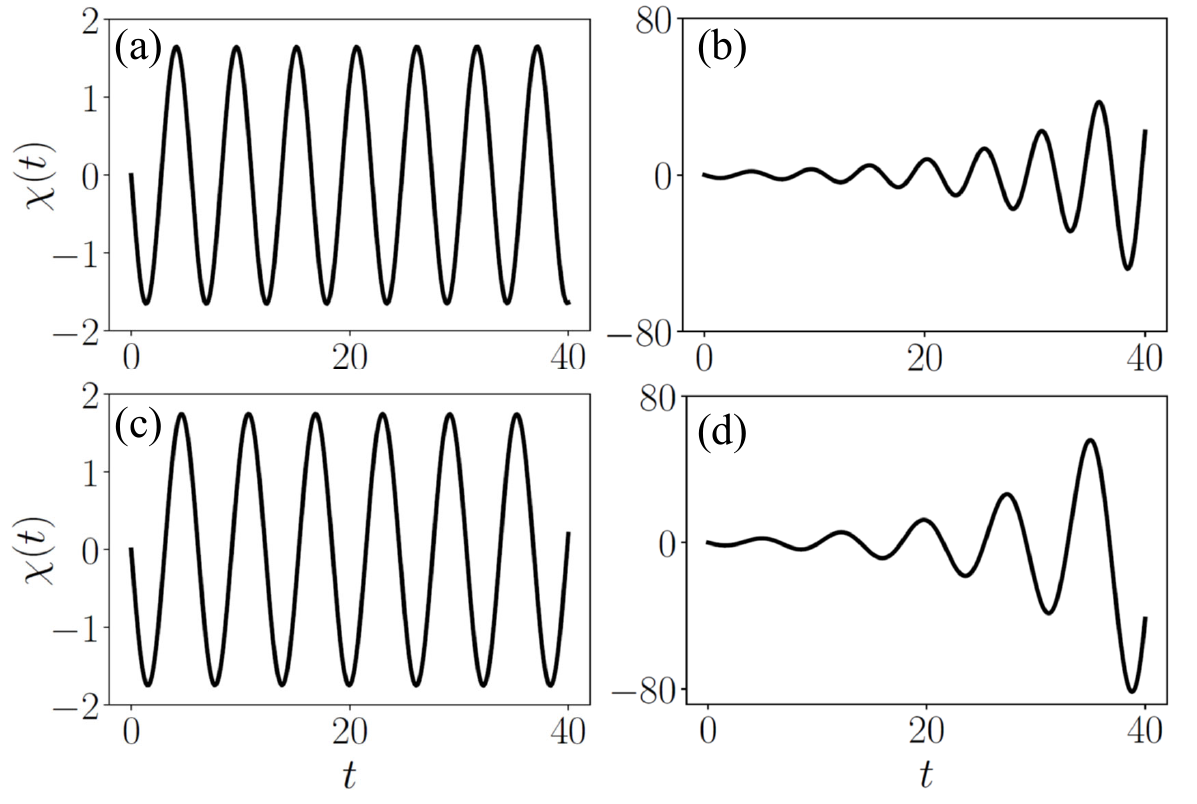}
\caption{Time-dependent linear response of the total polarization of two interacting qubits biased at $\epsilon/\Delta=0.2$ for different values of $\gamma=0$ (a),(c) and $\gamma/\Delta=0.2$ (b),(d). The interaction strength $g$ was chosen as $g/\Delta=-0.1$ (a),(b); $g/\Delta=0.1$ (c),(d).}
\label{fig:linear-response_e_02_G02}
\end{figure}

Fixing the interaction strength $g=\pm 0.1$ in the region where such specific $PT$-symmetry broken phase can be observed we compute the dynamic susceptibility of two biased interacting qubits, $\chi_{2Q}(t)$. In Fig. \ref{fig:linear-response_e_02_G02} the $\chi_{2Q}(t)$ is presented for two values of $\gamma$: $\gamma=0$ and $\gamma/\Delta=0.2$. If in the Hermitian case ($\gamma=0$) stable undamped quantum oscillations are found (see Fig. \ref{fig:linear-response_e_02_G02}a,b), the two \textit{PT}-symmetric biased interacting qubits demonstrate the \textit{amplified } oscillations (see Fig. \ref{fig:linear-response_e_02_G02}c,d) related to the ac induced transitions indicated by arrows in Fig. \ref{fig:spectrum_PD_e_02}.

\section{Conclusion}\label{chap6}
In conclusion, we have studied in detail the quantum dynamics of qubits systems whose behavior is governed by the \textit{PT}-symmetric non-Hermitian Hamiltonian. In particular, extending the famous Kubo linear response theory to pseudo-Hermitian Hamiltonian (pHH) systems we derive the general expressions for the temporal quantum-mechanical correlation function $C(t)$ and the time-dependent dynamic susceptibility $\chi(t) \propto \text{Im}~C(t)$  (see Eq. (\ref{eq: Correlationfunction})), determining the response of pHH systems to a small time-dependent perturbation, and identify various transitions between different states. Their amplitudes, however, are subjects of further normalization by $\langle \Psi(t)|\hat{\mathds{1}}|\Psi(t) \rangle_0$.

Next we have applied the elaborated generic theory to a study of the quantum dynamics of two $PT$-symmetric quantum systems: a single qubit and two interacting qubits. In such systems the $PT$-symmetry was introduced in a standard way through the gain and loss exactly compensating each other.
For both systems we observed the variety of \textit{PT}-symmetry unbroken and broken quantum phases, identify the quantum phase transitions between them as the parameters of gain/loss $\gamma$ and interaction strength $g$ vary. We derive a complete $\gamma-g$ phase diagram for two $PT$-symmetric interacting qubits. A peculiar result we obtain is that for two biased \textit{PT}-symmetric qubits the excited eigenstates show the \textit{ PT}-symmetry broken behavior even in the absence of the interaction between the qubits. 

The quantum dynamics of a $PT$-symmetric chain of qubits was characterized by the dynamic susceptibility of the total polarization of qubits, $\chi(t)$. The $\chi(t)$ displays oscillations and corresponding Fourier transform $\chi(\omega)$ shows a number of resonances that are the fingerprints of ac induced allowed transitions between the eigenstates.  Analyzing numerically calculated $\chi(t)$ we identify various transitions between the ground state and excited states for two $PT$-symmetric interacting qubits (indicated arrows in Figs. \ref{fig:spectrum_PD_e_0}b and  \ref{fig:spectrum_PD_e_02}b). 

The dynamic susceptibility of the total qubits polarization demonstrates a few generic and interesting features. In particular, for the $PT$-symmetry preserved phases the $\chi(t)$ dependence shows undamped quantum oscillations with the frequency that shifts to low values as the parameter $\gamma$ increases. Moreover, the transition between the \textit{PT}-symmetry broken ground state and unbroken excited state results in a  highly damped quantum oscillations in $\chi(t)$ dependence (see Fig. \ref{fig:linear-response_e_0_G0}h). Furthermore, for two biased interacting qubits we observe the opposite behavior, i.e. the \textit{amplification} of oscillations in $\chi(t)$ dependence, related to the ac induced transitions between the $PT$-symmetry unbroken ground state and broken excited state (see Fig. \ref{fig:linear-response_e_02_G02}d).
We anticipate that these peculiar features of $PT$-symmetry quantum systems can be observed in spectroscopic experiments with superconducting qubits. Our results can also be straightforwardly generalized to a long chain of qubits.

\textbf{Acknowledgements}
We thank B. Dora and V. Meden for fruitful discussions. We acknowledge the financial support through the European Union’s Horizon 2020 research and innovation program under grant agreement No 863313 'Supergalax'.\\

\appendix

\section{Matrix elements of the pseudo-Hermitian operators of a single qubit: \textit{PT}-symmetry preserved regime }\label{chapA}

In this Appendix we provide further details on the calculations of  the matrix elements of the pseudo-Hermitian operators for a single qubit system. 
By making use of the expression (\ref{SQ-etaoperator}) we obtain the operator $\hat \eta^{-1}$ as
\begin{align}
    \hat{\eta}^{-1} = \frac{\Delta^2}{E^2}\begin{pmatrix}1 & \mathrm{i}\frac{\gamma}{\Delta}\\ -\mathrm{i}\frac{\gamma}{\Delta} & 1\end{pmatrix}.\label{eqa1}
\end{align}
With that the pseudo-Hermitian operator $\hat \Sigma^z=\hat \eta^{-1} \sigma^z$ can be found as
\begin{align}
    \hat{\Sigma}^z =
    \frac{\Delta^2}{E^2}\begin{pmatrix}1 & \text{-}\mathrm{i}\frac{\gamma}{\Delta} \\ \text{-}\mathrm{i}\frac{\gamma}{\Delta} & -1 \end{pmatrix}.\label{eqa2}
\end{align}

In the PT-symmetric unbroken regime the corresponding matrix elements in the bi-orthogonal eigenbasis are
\begin{align}
    \langle R_{+}|\hat{\Sigma}^{z}|R_{+}\rangle &= 0\label{eqa3}\\
    \langle R_{-}|\hat{\Sigma}^{z}|R_{-}\rangle &= 0\label{eqa4}\\
    \langle R_{+}|\hat{\Sigma}^{z}|R_{-}\rangle &= \frac{\Delta^2}{E(E+\mathrm{i}\gamma)}\label{eqa5}\\
    \langle R_{-}|\hat{\Sigma}^{z}|R_{+}\rangle &= \frac{\Delta^2}{E(E-\mathrm{i}\gamma)}\label{eqa6}.
\end{align}
Using (\ref{eqa3})-(\ref{eqa6}) and the definition (\ref{eq: TDoperators}) of the time-dependent operator $\hat \Sigma^z(t)$ we obtain the  matrix element, which is necessary for the calculation of observable $\bar \sigma(t)$ as
$$
\langle R_{-}|\hat \eta \hat{\Sigma}^{z}(t)|R_{+}\rangle = \langle R_{-}|L_{-}\rangle\langle L_{-}|R_{-}\rangle \cdot \exp(-iEt/\hbar) \cdot
$$
\begin{equation}
\cdot \langle R_{-}|\hat{\Sigma}^{z}|R_{+}\rangle \cdot \exp(-iEt/\hbar) =
\frac{\Delta^2}{E(E-i\gamma)}\exp(-2iEt/\hbar)\label{eqa7}
\end{equation}
Here, we take into account the normalization condition, i.e. $\langle L_{\pm}|R_{\pm} \rangle =\delta_{\pm,\pm}$. 
Similarly, we obtain
$$
\langle R_{+}|\hat \eta \hat{\Sigma}^{z}(t)|R_{-}\rangle = \langle R_{+}|L_{+}\rangle\langle L_{+}|R_{+}\rangle \cdot \exp(iEt/\hbar) \cdot
$$
\begin{equation}
\cdot \langle R_+|\hat{\Sigma}^{z}|R_{-}\rangle \cdot \exp(iEt/\hbar) =
\frac{\Delta^2}{E(E+i\gamma)}\exp(2iEt/\hbar)\label{eqa8}
\end{equation}

\section{Matrix elements of the pseudo-Hermitian operators of a single qubit: \textit{PT}-symmetry broken regime}\label{chapB}

Here, we provide more details on the matrix elements of the pseudo-Hermitian operator $\hat \Sigma^z$ for a single qubit biased in the \textit{PT}-symmetry broken regime. Taking the eigenstates $|{R_\pm (L_{\pm})}\rangle$ as 
(\ref{eq:eigenvectorsBR}) and the explicit expression for the operator $\hat{\Sigma}^{z}$, i.e. Eq.  (\ref{SQ-Sigma}), we obtain the matrix elements of the operator $\hat{\Sigma}^{z}$ in the bi-orthogonal eigenbasis as
\begin{align}
    \langle R_{+}|\hat{\Sigma}_{z}|R_{+}\rangle &= 0\label{eqb3}\\
    \langle R_{-}|\hat{\Sigma}_{z}|R_{-}\rangle &= 0\label{eqb4}\\
    \langle R_{-}|\hat{\Sigma}_{z}|R_{+}\rangle &= \frac{\Delta^2}{\tilde{E}(\tilde{E}-\gamma)}\label{eqb5}\\
    \langle R_{+}|\hat{\Sigma}_{z}|R_{-}\rangle &= \frac{\Delta^2}{\tilde{E}(\tilde{E}+\gamma)}\label{eqb6}.
\end{align}
Similarly to the Appendix A we obtain
$$
\langle R_{-}|\hat \eta \hat{\Sigma}^{z}(t)|R_{-}\rangle = \langle R_{-}|L_{-}\rangle\langle L_{+}|R_{+}\rangle \cdot \exp(-\tilde{E}t/\hbar) \cdot
$$
\begin{equation}
\cdot \langle R_{+}|\hat{\Sigma}^{z}|R_{-}\rangle \cdot \exp(-\tilde{E}t/\hbar) =
\frac{\Delta^2}{\tilde{E}(\tilde{E}+\gamma)}\exp(-2\tilde{E}t/\hbar)\label{eqb7}
\end{equation}
and 
$$
\langle R_{+}|\hat \eta \hat{\Sigma}^{z}(t)|R_{+}\rangle = \langle R_{+}|L_{+}\rangle\langle L_{-}|R_{-}\rangle \cdot \exp(Et/\hbar) \cdot
$$
\begin{equation}
\cdot \langle R_{-}|\hat{\Sigma}^{z}|R_{+}\rangle \cdot \exp(Et/\hbar) =
\frac{\Delta^2}{\tilde{E}(\tilde{E}-\gamma)}\exp(2Et/\hbar)\label{eqb8}
\end{equation}



\bibliography{bibliography}
\end{document}